\documentclass{article}
\usepackage{ijcai18}

\usepackage{times}
\usepackage{xcolor}
\usepackage{soul}
\usepackage[utf8]{inputenc}
\usepackage[small]{caption}
\usepackage{amsmath}
\usepackage{mathtools}  
\usepackage{amsfonts}
\usepackage{hyperref}
\usepackage{enumitem}

\usepackage[font=small,labelfont=bf,tableposition=top]{caption}

\title{ Discrete Factorization Machines for Fast Feature-based Recommendation
\thanks{This work is accepted by IJCAI 2018.}
}
\author{
Han Liu$^1$,
Xiangnan He$^2$,
Fuli Feng$^2$,
Liqiang Nie$^1$,
Rui Liu$^3$,
Hanwang Zhang$^4$
\\
$^1$School of Computer Science and Technology, Shandong University\\
$^2$School of Computing, National University of Singapore\\
$^3$University of Electronic Science and Technology of China\\
$^4$School of Computer Science and Engineering, Nanyang Technological University\\
\{hanliu.sdu,
xiangnanhe,
fulifeng93,
nieliqiang,
ruiliu011,
hanwangzhang\}@gmail.com
}
\begin{document}

\maketitle
\begin{abstract}
User and item features of side information are crucial for accurate recommendation. However, the large number of feature dimensions, \textit{e.g.}, usually larger than $10^7$, results in expensive storage and computational cost. This prohibits fast recommendation especially on mobile applications where the computational resource is very limited. In this paper, we develop a generic feature-based recommendation model, called \textit{Discrete Factorization Machine} (DFM), for fast and accurate recommendation. DFM binarizes the real-valued model parameters (\textit{e.g.}, float32) of every feature embedding into binary codes (\textit{e.g.}, boolean), and thus supports efficient storage and fast user-item score computation. To avoid the severe quantization loss of the binarization, we propose a convergent updating rule that resolves the challenging discrete optimization of DFM.   Through extensive experiments on two real-world datasets,
  we show that 1) DFM consistently outperforms state-of-the-art binarized recommendation models, and 2) DFM shows very competitive performance compared to its real-valued version (FM), demonstrating the minimized quantization loss.
\end{abstract}

\section{Introduction}
Recommendation is ubiquitous in today's cyber-world --- almost every one of your Web activities can be viewed as a recommendation, such as news or music feeds, car or restaurant booking, and online shopping. Therefore, accurate recommender system is not only essential for the quality of service, but also the profit of the service provider. One such system should exploit the rich side information beyond user-item interactions, such as content-based (\textit{e.g.}, user attributes~\cite{Silkroad} and product image features~\cite{Yu:2018:ACR}), context-based (\textit{e.g.}, 
where and when a purchase is made~\cite{rendle2011fast,NFM}), and session-based (\textit{e.g.}, the recent browsing history of users~\cite{Li:2017:NAS:3132847.3132926,iCD}). However, existing collaborative filtering (CF) based systems merely rely on user and item features (\textit{e.g.}, matrix factorization based~\cite{fastMF} and the recently proposed neural collaborative filtering methods~\cite{NCF,bai2017neural}), which are far from sufficient to capture the complex decision psychology of the setting and the mood of a user behavior~\cite{ACF}.

Factorization Machine (FM)~\cite{Rendle2011Factorization} is one of the prevalent feature-based recommendation model that leverages rich features of users and items for accurate recommendation. FM can incorporate any side features by concatenating them into a high-dimensional and sparse feature vector. The key advantage of it is to learn $k$-dimensional latent vectors , \textit{i.e.}, the embedding parameters $\mathbf{V}\in\mathbb{R}^{k\times n}$, for all the $n$ feature dimensions. They are then used to model pairwise interactions between features in the embedding space. However, since $n$ is large (\textit{e.g.} practical recommender systems typically need to deal with over millions of items and other features where $n$ is at least $10^7$~\cite{Wang:2018:PFD}), it is impossible on-device storage of $\mathbf{V}$. 
Moreover, it requires large-scale multiplications of the feature interaction $\mathbf{v}^T_i\mathbf{v}_j$ for user-item score, even linear time-complexity is prohibitively slow for float operations. Therefore, existing FM framework is not suitable for fast recommendation, especially for mobile users.

In this paper, we propose a novel feature-based recommendation framework, named \textit{Discrete Factorization Machine} (DFM), for fast recommendation. In a nutshell, DFM replaces the real-valued FM parameters $\mathbf{V}$ by binary-valued $\mathbf{B}\in\{\pm 1\}^{k\times n}$. In this way, we can easily store a  bit matrix and perform XOR bit operations instead of float multiplications, making fast recommendation on-the-fly possible. However, it is well-known that the binarization of real-valued parameters will lead to significant performance drop due to the quantization loss~\cite{Zhang2016Discrete}. To this end, we propose to directly optimize the binary parameters in an end-to-end fashion, which is fundamentally different from the widely adopted two-stage approach that first learns real-valued parameters and then applies round-off binarization~\cite{Zhang2014Preference}. Our algorithm jointly optimize the two challenging objectives: 1) to tailor the binary codes $\mathbf{B}$ to fit the original loss function of FM, and 2) imposing the binary constraint that is balanced and decorrelated, to encode compact information. In particular, we develop an alternating optimization algorithm to iteratively solve the mixed-integer programming problems. We evaluate DFM on two real-world datasets Yelp and Amazon, the results demonstrate that 1) DFM consistently outperforms state-of-the-art binarized recommendation models, and 2) DFM shows very competitive performance compared to its real-valued version (FM), demonstrating the minimized quantization loss. 

Our contributions are summarized as follows:
\begin{itemize}[leftmargin=*]
\item We propose to binarize FM, a dominant feature-based recommender model, to enable fast recommendation. To our knowledge, this is the first generic solution for fast recommendation that learns a binary embedding for each feature.
\item We develop an efficient algorithm to address the challenging optimization problem of DFM that involves discrete, balanced, and de-correlated constraints.
\item Through extensive experiments on two real-world datasets,
we demonstrate that DFM outperforms state-of-the-art hash-based recommendation algorithms.
\end{itemize}

\section{Related Work}
We first review efficient recommendation algorithms
using latent factor models,
and then discuss recent advance in discrete hashing techniques.

\subsection{Efficient Recommendation}
As pioneer work,
\cite{Das2007Google} used Locality-Sensitive Hashing (LSH) \cite{Gionis1999Similarity}
to generate hash codes for Google new users based on their item-sharing history similarity.
Following the work,
\cite{Karatzoglou2010Collaborative} applied random projection for mapping learned user-item latent factors from traditional CF into the Hamming space to acquire hash codes for users and items.
Similar to the idea of projection,
\cite{Zhou2012Learning} generate binary code from rotated continuous user-item latent factors by running ITQ \cite{Gong2011Iterative}.
In order to derive more compact binary codes,
\cite{Liu2014Collaborative} imposed the de-correlation constraint of different binary codes on continuous user-item latent factors and then rounded them to produce binary codes.
However, \cite{Zhang2014Preference}
argued that hashing only preserves similarity between user and item rather than inner product based preference,
so subsequent hashing may harm the accuracy of preference predictions,
thus they imposed a Constant Feature Norm(CFN) constraint on user-item continuous latent factors,
and then quantized similarities by respectively thresholding their magnitudes and phases.

The aforementioned work can be easily summarized as two independents stages:
relaxed user-item latent factors learning with some specific constraints and binary quantization.
However, such a two-stage relaxation is well-known to suffer from a large quantization loss according to \cite{Zhang2016Discrete}.

\subsection{Binary Codes Learning}
Direct binary code learning by discrete optimization --- is becoming popular recently in order to decrease quantization loss aforementioned. 
Supervised hashing~\cite{Luo:2018} improve on joint optimizations of quantization losses and intrinsic objective functions, whose significant performance gain over the above two-stage approaches.

In the recommendation area, \cite{Zhang2016Discrete} is the first work that proposes to learn binary codes for users and items by directly optimizing the recommendation task. The proposed method \textit{Discrete Collaborative Filtering} (DCF) demonstrates superior performance over aforementioned two-stage efficient recommendation methods. To learn informative and compact codes, DCF proposes to enforce balanced and de-correlated constraints on the discrete optimization. Despite its effectiveness, DCF models only user-item interactions and cannot be trivially extended to incorporate side features. As such, it suffers from the cold-start problem and can not be used as a generic recommendation solution, e.g., for context-aware~\cite{Rendle2011Factorization} and session-based recommendation~\cite{iCD}. 
Same as the relationship between FM and MF, our DFM method can be seen as a generalization of DCF that can be used for generic feature-based recommendation. Specifically, feeding only ID features of users and items to DFM will recover the DCF method. In addition, our DFM can learn binary codes for each feature, allowing it to be used for resource-limited recommendation scenarios, such as context-aware recommendation in mobile devices. This binary representation learning approach for feature-based recommendation, to the best of knowledge, has never been developed before. 

The work that is most relevant with this paper is \cite{Lian2017Discrete}, which develops a discrete optimization algorithm named \textit{Discrete Content-aware Matrix Factorization} (DCMF), to learn binary codes for users and items at the presence of their respective content information.
It is worth noting that DCMF can only learn binary codes for each user ID and item ID, rather than their content features. Since its prediction model is still MF (\textit{i.e.,}, the dot product of user codes and item codes only), it is rather limited in leveraging side features for accurate recommendation.
As such, DCMF only demonstrates minor improvements over DCF for feature-based collaborative recommendation~(\textit{cf.} Figure 2(a) for their original paper). 
Going beyond learning user codes and item codes, our DFM can learn codes for any side feature and model the pairwise interactions between feature codes. As such, our method has much stronger representation ability than DCMF, demonstrating significant improvements over DCMF in feature-based collaborative recommendation. 
\section{Preliminaries}
We use bold uppercase and lowercase letters as matrices and vectors, respectively.
In particular, we use $\mathbf{a}_i$ as the $a$-th column vector of matrix $\mathbf{A}$.
We denote ${\|\cdot\|}_F$ as the Frobenius norm of a matrix and $\text{tr}(\cdot)$ as the matrix 
trace.
We denote $\text{sgn}(\cdot):\mathbb{R}\rightarrow \{\pm 1\}$
as the round-off function, \textit{i.e.}, $\text{sgn}(x) = +1$ if $x\geq 0$ and $\text{sgn}(x) = -1$ otherwise.

Factorization Machine (FM) is essentially a score prediction function for a (user, item) pair feature $\mathbf{x}$:
\begin{equation}
\label{eq:fm}
\small
\text{FM}(\mathbf{x}):= w_{0}+\sum\limits_{i=1}^{n} w_i x_i+
\sum\limits_{i=1}^{n}\sum\limits_{j=i+1}^{n}\langle \mathbf{v}_i,\mathbf{v}_j\rangle x_i x_j,
\end{equation}
where $\mathbf{x}\in\mathbb{R}^n$ is a high-dimensional feature representation of the rich side-information, concatenated by one-hot user ID and item ID, user and item content features, location features, \textit{etc}. $\mathbf{w}\in\mathbb{R}^n$ is the model bias parameter: $w_o$ is the global bias and $w_i$ is the feature bias. $\mathbf{V}\in\mathbb{R}^{k\times n}$ is the latent feature vector, and every $\langle \mathbf{v}_i,\mathbf{v}_j\rangle$ models the interaction between the $i$-th and $j$-th feature dimensions. Therefore, $\mathbf{V}$ is the key reason why FM is an effective feature-based recommendation model, as it captures the rich side-information interaction. However, on-the-fly storing $\mathbf{V}$ and computing $\langle \mathbf{v}_i,\mathbf{v}_j\rangle$ are prohibitively expensive when $n$ is large. For example, a practical recommender system for  Yelp\footnote{\href{https://www.yelp.ca/dataset}{https://www.yelp.ca/dataset}} needs to provide recommendation for over $1,300,000$ users with about $174,000$ business, which have more than $1,200,000$ attributes (here, $n=1,300,000+174,000+1,200,000=2,674,000$). 

To this end, we want to use binary codes $\mathbf{B}\in\{\pm 1\}^{k\times n}$ instead of $\mathbf{V}$, to formulated our proposed framework: Discrete Factorization Machines (DFM):
\begin{equation}\label{eq:dfm}\small
\text{DFM}(\mathbf{x}):= w_{0}+\sum\limits_{i=1}^{n} w_i x_i+
\sum\limits_{i=1}^{n}\sum\limits_{j=i+1}^{n}\langle \mathbf{b}_i,\mathbf{b}_j\rangle x_i x_j.
\end{equation}
However, directly obtain $\mathbf{B} = \textrm{sgn}(\mathbf{V})$ will lead to large quantization loss and thus degrade the recommendation accuracy significantly~\cite{Zhang2016Discrete}. In the next section, we will introduce our proposed DFM learning model and discrete optimization that tackles the quantization loss.

\section{Discrete Factorization Machines}
We first present the learning objective of DFM and then elaborate the optimization process of DFM, which is the key technical difficulty of the paper. At last, we shed some lights on model initialization, which is known to have a large impact on a discrete model.

\subsection{Model Formulation}
Given a training pair $(\mathbf{x},y)\in\mathcal{V}$, where $y$ is the groundtruth score of feature vector $\textbf{x}$ and $\mathcal{V}$ denotes the set of all training instances, the problem of DFM is formulated as:
\begin{align}\small
&\mathop{\arg\min}\limits_{w_0,\mathbf{w},\mathbf{B}}
     \sum\limits_{(\mathbf{x},y)\in \mathcal{V}}
    (y-w_{0}-\sum\limits_{i=1}^{n} w_i x_i
    -\sum\limits_{i=1}^{n}\sum\limits_{j=i+1}^{n}\langle \mathbf{b}_i,\mathbf{b}_j\rangle x_i x_j)^2 \notag \\
    &+ \alpha\sum\limits_{i=1}^{n} w_i^2,
    \text{s.t.}\ \mathbf{B} \in\{\pm1\}^{k\times n},\ \underbrace{\mathbf{B}\mathbf{1} = \mathbf{0}}_{\text{Balance}},\
    \underbrace{
        \mathbf{B}\mathbf{B}^T = n\mathbf{I}
    }_{\text{De-correlation}}
\label{eq:obj}
\end{align}
Due to the discrete constraint in DFM,
the regularization ${\|\mathbf{B}\|}_F^2$ becomes an constant and hence is removed.
Additionally, DFM imposes balanced and de-correlated constraints on the binary codes in order to maximize the information each bit carries and to make binary codes compact \cite{Zhou2012Learning}. 
However, optimizing the objective function in Eq.(\ref{eq:obj}) is a highly challenging task, since it is
generally NP-hard. To be specific, finding the global optimum solution needs to involve $\mathcal{O}(2^{kn})$ combinatorial search for the binary codes~\cite{Stad2001Some}.

Next, we introduce a new learning objective that allows DFM to be solved in a computationally tractable way. The basic idea is to soften the balanced and de-correlated constraints.
To achieve this, let us first introduce a delegate continuous variable $\mathbf{D}\in\mathcal{D}$,
where $\mathcal{D}=\{\mathbf{D}\in\mathbb{R}^{k\times n}|\mathbf{D}\mathbf{1} = \mathbf{0},\mathbf{D}\mathbf{D}^T = n\mathbf{I}\}$.
Then the balanced and de-correlated constraints can be softened by
$\min_{D\in\mathcal{D}}\|\mathbf{B}-\mathbf{D}\|_F$.
As such, we can get the softened learning objective for DFM as:
\begin{align}\small\label{eq:softobj}
\mathop{\arg\min}\limits_{w_0,\mathbf{w},\mathbf{B}}
    \sum\limits_{(\mathbf{x},y)\in \mathcal{V}}
    &(y-w_{0}-\sum\limits_{i=1}^{n} w_i x_i-\sum\limits_{i=1}^{n}\sum\limits_{j=i+1}^{n}\langle \mathbf{b}_i,\mathbf{b}_j\rangle x_i x_j)^2
    \notag \\
    &+ \alpha\sum\limits_{i=1}^{n} w_i^2 - 2\beta tr(\mathbf{B}^T\mathbf{D}),
    \\ \notag
    \text{s.t.}\ &\mathbf{D}\mathbf{1} = \mathbf{0},\mathbf{D}\mathbf{D}^T = n\mathbf{I},\mathbf{B}\in\{\pm 1\}^{k\times n},
\end{align}
where we use $2tr(\mathbf{B}^T\mathbf{D})$ instead of $\|\mathbf{B}-\mathbf{D}\|_F$ for the ease of optimization (note that the two terms are identical since $\mathbf{B}^T\mathbf{B}$ and $\mathbf{D}^T\mathbf{D}$ are constant). $\beta$ is tunable hyperparameter controlling the strength of the softened de-correlation constraint.
As the above Eq.(\ref{eq:softobj}) allows a certain discrepancy between $\mathbf{B}$ and $\mathbf{D}$,
it makes the binarized optimization problem computationally tractable.
Note that if there are feasible solution in Eq.(\ref{eq:obj}),
we can impose a very large $\beta$ to enforce $\mathbf{B}$ to be close to $\mathbf{D}$.

The above Eq.(\ref{eq:softobj}) presents the objective function to be optimized for DFM.
It is worth noting that we do not discard the discrete constraint and we still perform a direct optimization on discrete $\mathbf{B}$.
Furthermore, through joint optimization for the binary codes and the delegate real variables,
we can obtain nearly balanced and uncorrelated binary codes.
Next, we introduce an efficient solution to solve the mixed-integer optimization problem in Eq.(\ref{eq:softobj}).

\subsection{Optimization}
We employ alternating optimization strategy~\cite{liu2017pami} to solve the problem. Specifically, we alternatively solve three subproblems for DFM model in Eq.(\ref{eq:softobj}),
taking turns to update each of $\mathbf{B}$, $\mathbf{D}$, $\mathbf{w}$,
given others fixed. Next we elaborate on how to solve each of the subproblems. 

\noindent $\mathbf{B}$\textbf{-subproblem}.\quad
In this subproblem, we aim to optimize $\mathbf{B}$ with  fixed $\mathbf{D}$ and $\mathbf{w}$. 
To achieve this, we can update $\mathbf{B}$ by updating each vector $\mathbf{b}_r$ according to
\begin{equation*}
\begin{aligned}\small
&\mathop{\arg\min}\limits_{\mathbf{b}_r\in\{\pm 1\}^k}
    \mathbf{b}_r^T\mathbf{U}
    (\sum\limits_{\mathcal{V}_r}x_r^2 \hat{\mathbf{x}} \hat{\mathbf{x}} ^T)
    \mathbf{U}^T\mathbf{b}_r
    -2(\sum\limits_{\mathcal{V}_r}x_r\psi  \hat{\mathbf{x}} ^T )\mathbf{U}^T \mathbf{b}_r    \\
    &-2\beta \mathbf{d}_r^T\mathbf{b}_r,\ 
    \text{where}\ \psi =
    y-w_0 -
    \textbf{w}^T \textbf{x}  -
    \sum\limits_{i=1}^{n-1}\sum\limits_{j=i+1}^{n-1}\langle \mathbf{u}_i,\mathbf{u}_j\rangle \hat{x}_i \hat{x}_j
\end{aligned}
\end{equation*}
where $\mathcal{V}_r=\{(\mathbf{x},y)\in \mathcal{V}|x_r\neq 0\}$
is the training set for $\mathbf{r}$,
vector $\hat{\mathbf{x}}$ is equal to $\mathbf{x}$ excluding element $x_r$,
$\mathbf{U}$ excludes the column $\mathbf{b}_r$ of matrix $\mathbf{B}$,
and $\mathbf{u}_i$ is a column in $\mathbf{U}$.

Due to the discrete constraints, the optimization is generally NP-hard. To this end, we use Discrete Coordinate Descent (DCD)~\cite{Zhang2016Discrete} to take turns to update each bit of binary codes $\mathbf{b}_r$.
Denote $b_{rt}$ as the $t$-th bit of $\mathbf{b}_r$ and
$\mathbf{b}_{r\bar{t}}$ as the rest codes excluding $b_{rt}$,
DCD will update $b_{rt}$ by fixing $\mathbf{b}_{r\bar{t}}$.
Thus, we update $b_{rt}$ based on the following rule:
\begin{equation}\small
\begin{split}
&b_{rt}\leftarrow\text{sgn}\big( K(\hat{b}_{rt},b_{rt})\big),\\
\hat{b}_{rt}=\sum_{\mathcal{V}_r}
    &(x_r\psi-x_r^2\hat{\mathbf{x}} ^T\mathbf{Z}_{\bar{t}}\mathbf{b}_{r\bar{t}})
    \hat{\mathbf{x}}^T\mathbf{z}_t
    +\beta d_{rt}
\end{split}
\end{equation}
where
$\mathbf{Z}=\mathbf{U}^T$,
$\mathbf{z}_t$ is the $t$-th column of the matrix $\mathbf{Z}$
while $\mathbf{Z}_{\bar{t}}$ excludes the $t$-th column from $\mathbf{Z}$,
and $K(x,y)$ is a function that $K(x,y)=x$ if $x\neq 0$
and $K(x,y)=y$ otherwise.
Through this way, we can control that when $\hat{b}_{rt}=0$, we do not update $b_{rt}$. \vspace{+5pt}

\noindent $\mathbf{D}$\textbf{-subproblem}.\quad
When $\mathbf{B}$ and $\mathbf{w}$ are fixed in Eq.(\ref{eq:softobj}),
the optimization subproblem for $\mathbf{D}$ is:
\begin{equation}\label{eq:dsubp}\small
\mathop{\arg\max}\limits_{\mathbf{D}}tr(\mathbf{B}^T\mathbf{D}),
s.t.\ \mathbf{D}\mathbf{1}=\mathbf{0},
\mathbf{D}\mathbf{D}^T=m\mathbf{I}.
\end{equation}
It can be solved with the aid of a centering matrix
$\mathbf{J}=\mathbf{I}-\frac{1}{n}\mathbf{1}\mathbf{1}^T$.
Specifically, by Singular Value Decomposition (SVD),
we have $\mathbf{B}\mathbf{J}=\overline{\mathbf{B}}=\mathbf{P}\mathbf{\Sigma}\mathbf{Q}^T$,
where $\mathbf{P}\in\mathbb{R}^{k\times k'}$
and $\mathbf{Q}\in\mathbb{R}^{n\times k'}$
are left and right singular vectors corresponding to the $r' (\leq r)$
positive singular values in the diagonal matrix $\mathbf{\Sigma}$.
We first apply eigendecomposition for the small $k\times k$ matrix
$\overline{\mathbf{B}}\ \overline{\mathbf{B}}^T=
\begin{bmatrix}
\mathbf{P}&\widehat{\mathbf{P}}
\end{bmatrix}
\begin{bmatrix}
\mathbf{\Sigma}^2&\mathbf{0}\\ \mathbf{0}&\mathbf{0}
\end{bmatrix}
\begin{bmatrix}
\mathbf{P}&\widehat{\mathbf{P}}
\end{bmatrix}^T$,
where $\widehat{\mathbf{P}}$ are the eigenvectors of the zero eigenvalues.
Therefore, by the definition of SVD,
we have $\mathbf{Q}=\overline{\mathbf{B}}^T\mathbf{P}\mathbf{\Sigma}^{-1}$.
In order to satisfy the constraint $\mathbf{D}\mathbf{1}=0$,
we further obtain additional $\widehat{\mathbf{Q}}\in\mathbb{R}^{n\times(k-k')}$
by Gram-Schmidt orthogonalization based on
$\begin{bmatrix}
\mathbf{Q}&\mathbf{1}
\end{bmatrix}$. 
As such, we have $\widehat{\mathbf{Q}}^T\mathbf{1}=\mathbf{0}$.
Then we can get the closed-form update rule for the $\mathbf{D}$-subproblem in Eq.(\ref{eq:dsubp})
as:
\begin{equation}\small
\mathbf{D}\leftarrow\sqrt{n}
\begin{bmatrix}
\mathbf{P}&\widehat{\mathbf{P}}
\end{bmatrix}
\begin{bmatrix}
\mathbf{Q}&\widehat{\mathbf{Q}}
\end{bmatrix}^T
\end{equation}
\noindent $\mathbf{w}$\textbf{-subproblem}.\quad
When $\mathbf{B}$ and $\mathbf{D}$ are fixed in Eq.(\ref{eq:softobj}),
the subproblem is for optimizing $\mathbf{w}$ is:
\begin{equation}\small
\begin{split}
\mathop{\arg\min}\limits_{w_0,\mathbf{w}}
    &\sum\limits_{(\mathbf{x},y)\in\mathcal{V}}
    (\phi -w_{0}-\sum\limits_{i=1}^{n} w_i x_i)^2
    +\alpha\sum\limits_{i=1}^{n} w_i^2
    ,\\
    &\phi = y-\sum\limits_{i=1}^{n}\sum\limits_{j=i+1}^{n}\langle \mathbf{b}_i,\mathbf{b}_j\rangle x_i x_j.
\end{split}
\end{equation}
Since $\textbf{w}$ is a real-valued vector, it is the standard multivariate linear regression problem.
Thus we can use coordinate descent algorithm provided in the original FM~\cite{Rendle2011Factorization} to find the optimal value of $\mathbf{w}$ and the global bias $w_0$.

\subsection{Initialization}
Since DFM deals with mixed-integer non-convex optimization, the initialization of model parameters plays an important role for faster convergence and for finding better local optimum solution.
Here we suggest an efficient initialization strategy inspired by DCF~\cite{Zhang2016Discrete}. It first solves a relaxed optimization problem in Eq.(4) by discarding the discrete constraints as:
\begin{equation}
\begin{small}
\begin{aligned}
\label{eq:init}
\small
&\mathop{\arg\min}\limits_{w_0,\mathbf{w},\mathbf{V}}
    \sum\limits_{(\mathbf{x},y)\in \mathcal{V}}
    (y-w_{0}-\sum\limits_{i=1}^{n} w_i x_i-\sum\limits_{i=1}^{n}\sum\limits_{j=i+1}^{n}\langle \mathbf{v}_i,\mathbf{v}_j\rangle x_i x_j)^2\notag
    \\
    + &\alpha
    \sum\limits_{i=1}^{n} w_i^2 + \beta\|\mathbf{V}\|_F^2
    - 2\beta tr(\mathbf{V}^T\mathbf{D}), \text{s.t.}\ \mathbf{D}\mathbf{1} = \mathbf{0},\mathbf{D}\mathbf{D}^T = n\mathbf{I}\notag
\end{aligned}
\end{small}
\end{equation}
To solve the problem,
we can initialize real-valued $\mathbf{V}$ and $\mathbf{w}$ randomly
and find feasible initializations for $\mathbf{D}$
by solving $\mathbf{D}$-subproblem.
The optimization can be done alternatively by solving $\mathbf{V}$ by traditional FM,
solving $\mathbf{D}$ by $\mathbf{D}$-subproblem,
and solving $\mathbf{w}$ by gradient descent.
Assuming the solution is ($\mathbf{V}^\ast,\mathbf{D}^\ast,\mathbf{w}^\ast,w_0^\ast$),
we can then initialize the parameters in Eq.(\ref{eq:softobj}) as:
\begin{equation}
\mathbf{B}\leftarrow\text{sgn}(\mathbf{V}^\ast),
\mathbf{D}\leftarrow\mathbf{D}^\ast,
\mathbf{w}\leftarrow\mathbf{w}^\ast,
w_0\leftarrow w_0^\ast
\end{equation}

\section{Experiments}
As the key contribution of this work is the design of DFM for fast feature-based recommendation, we conduct experiments to answer the following research questions:
~\\
\noindent \textbf{RQ1}.\quad
How does DFM perform as compared to existing hash-based recommendation methods?
~\\
\noindent \textbf{RQ2}.\quad
How does the key hyper-parameter of DFM impact its recommendation performance?
~\\
\noindent \textbf{RQ3}.\quad
How efficient is DFM as compared to the real-valued version of FM?

\begin{figure*}[!tbh]
\centering
\includegraphics[scale = 1.5]{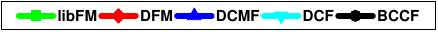}\\
\large\textbf{Yelp}\\
\vspace{-0.2cm}
\includegraphics[width=0.265\textwidth]{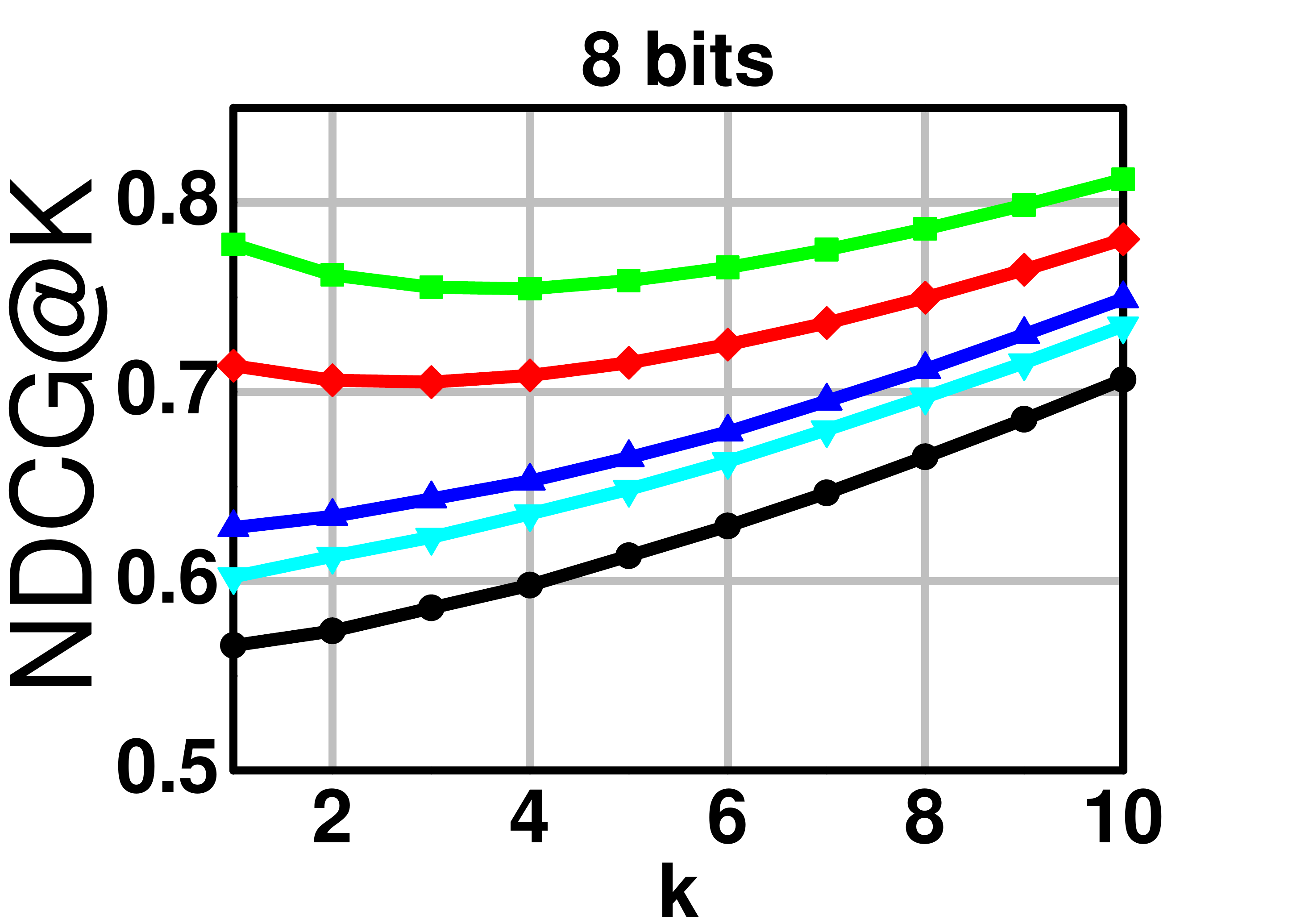}
\hspace{-0.24in}
\includegraphics[width=0.265\textwidth]{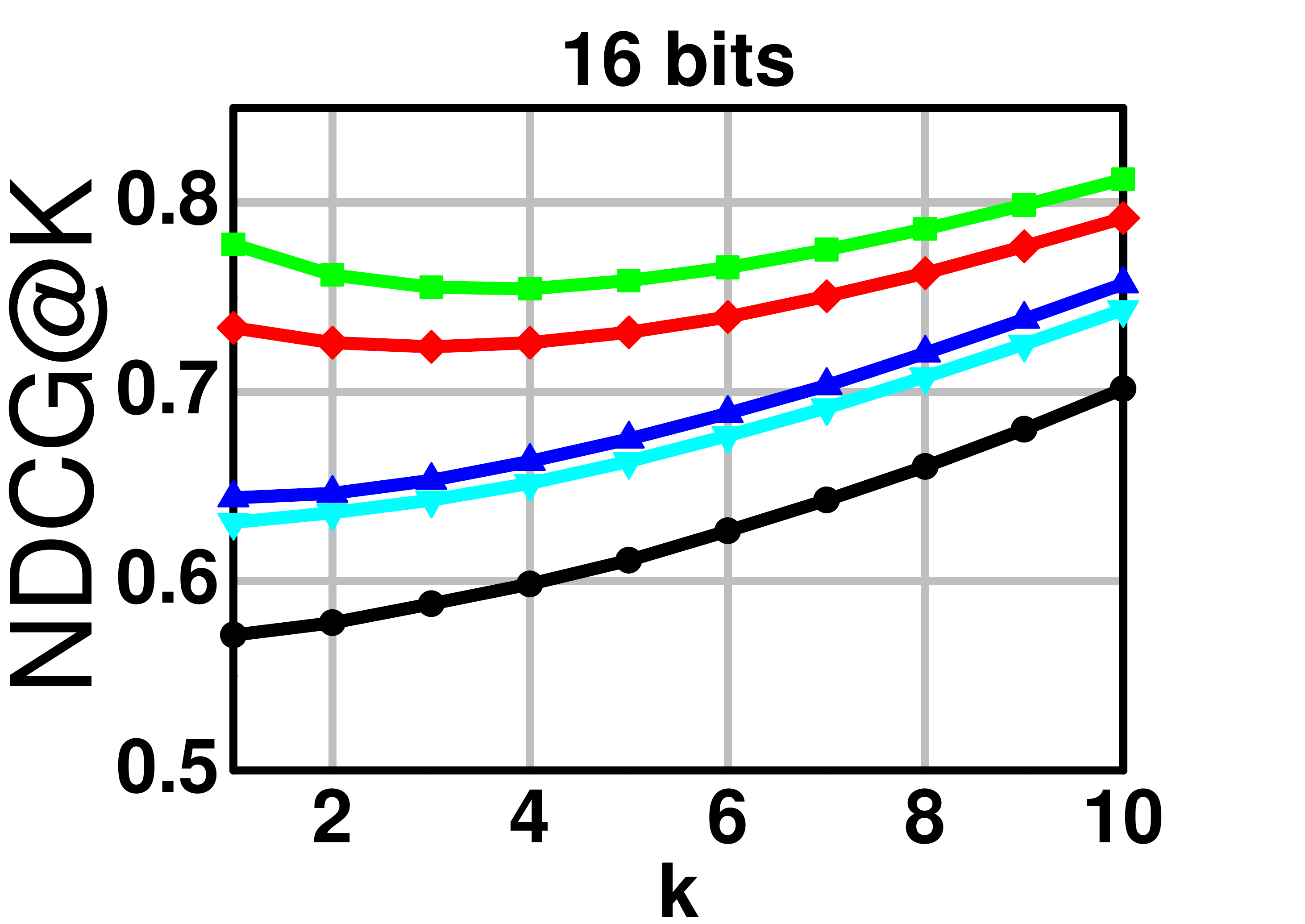}
\hspace{-0.24in}
\includegraphics[width=0.265\textwidth]{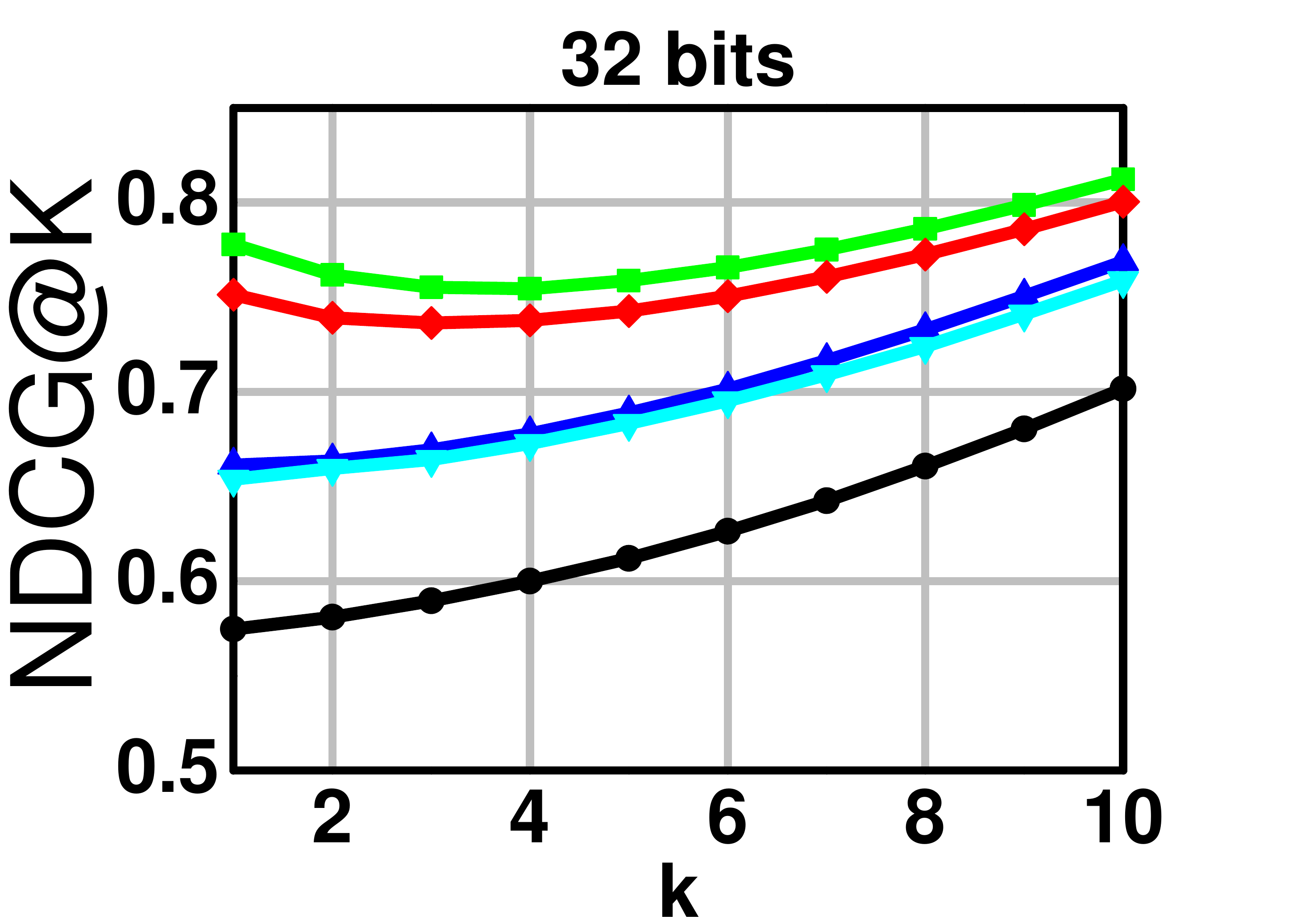}
\hspace{-0.24in}
\includegraphics[width=0.265\textwidth]{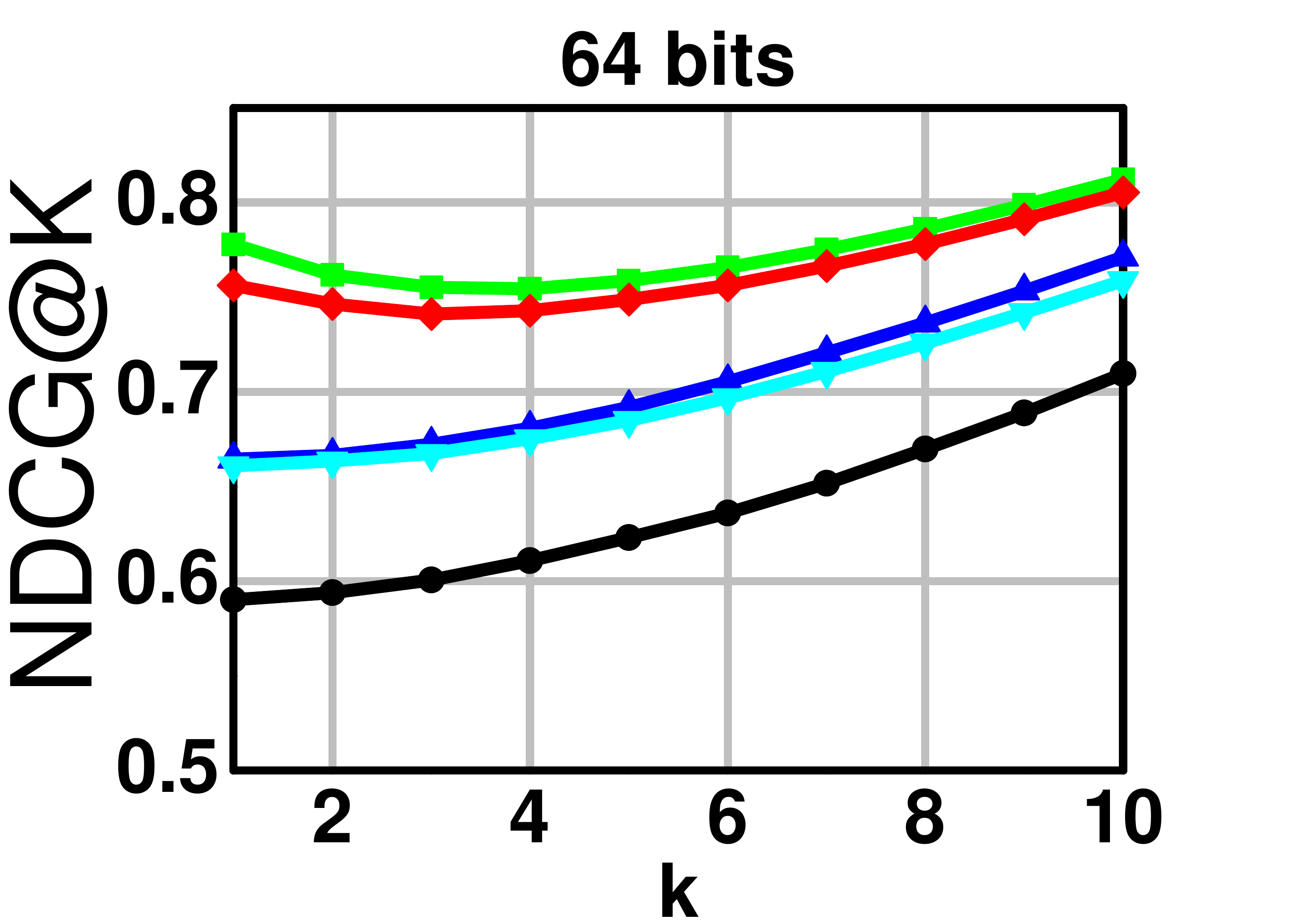}
\\
\vspace{-0.2cm}
\large\textbf{Amazon}\\
\includegraphics[width=0.265\textwidth]{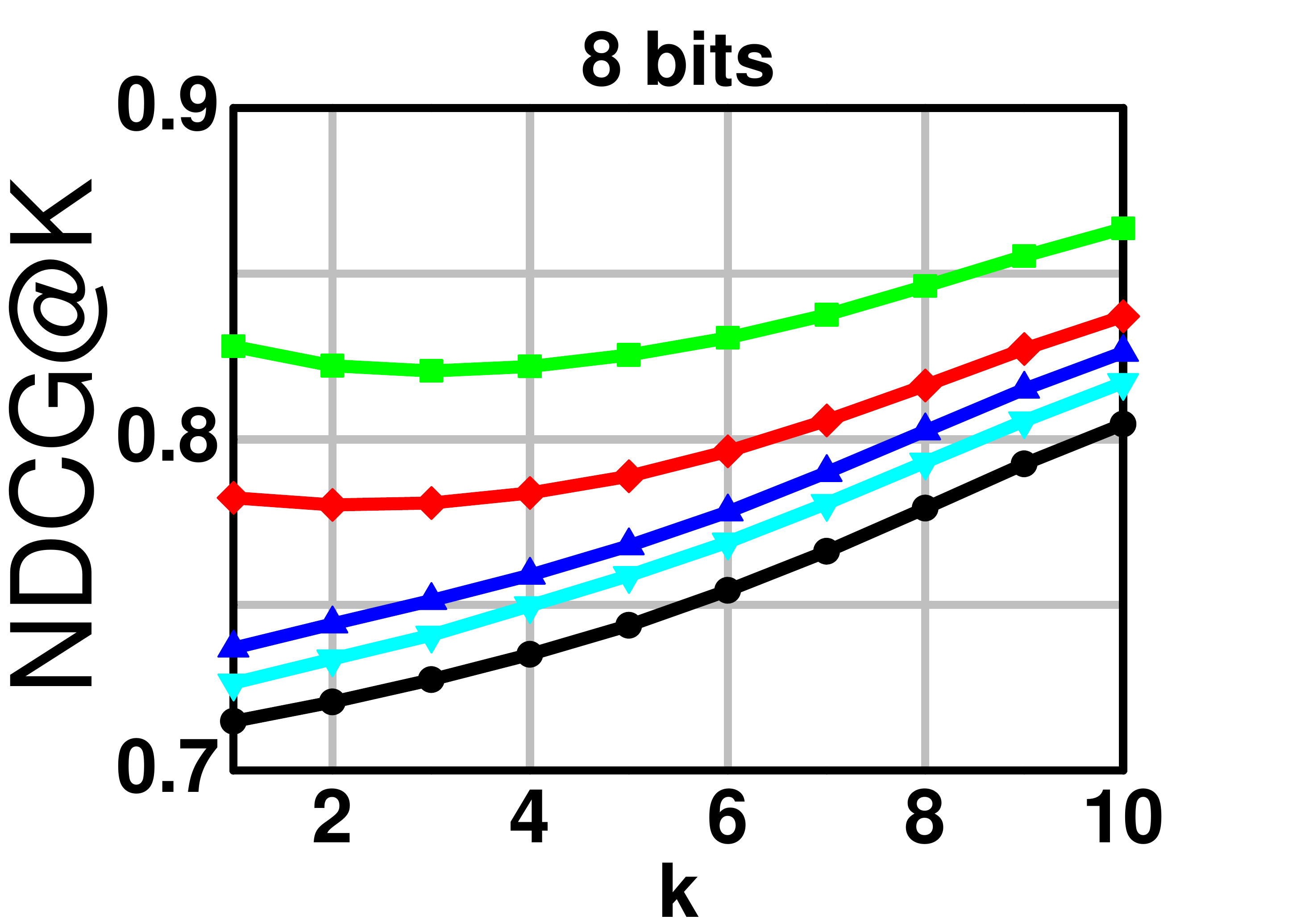}
\hspace{-0.24in}
\includegraphics[width=0.265\textwidth]{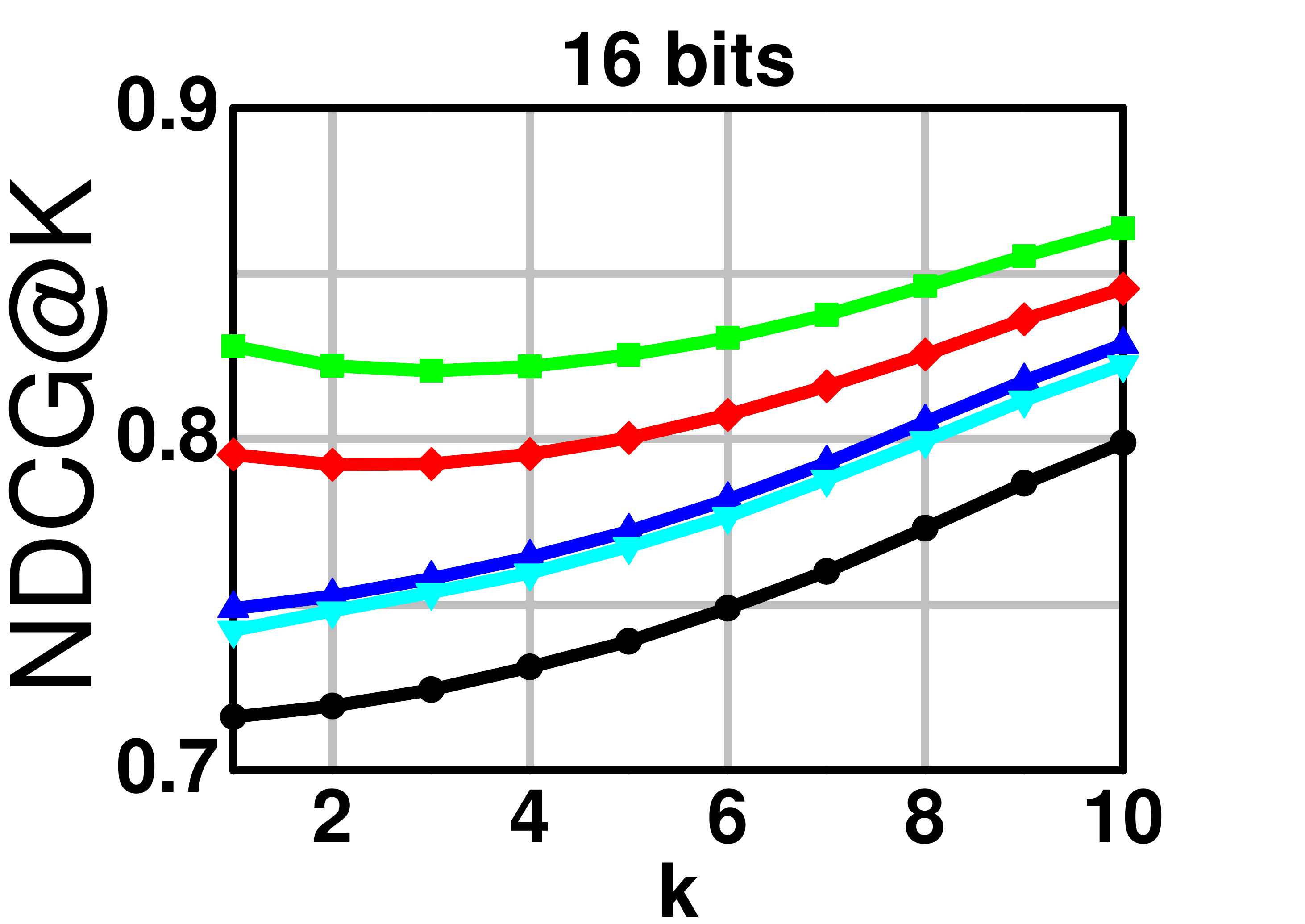}
\hspace{-0.24in}
\includegraphics[width=0.265\textwidth]{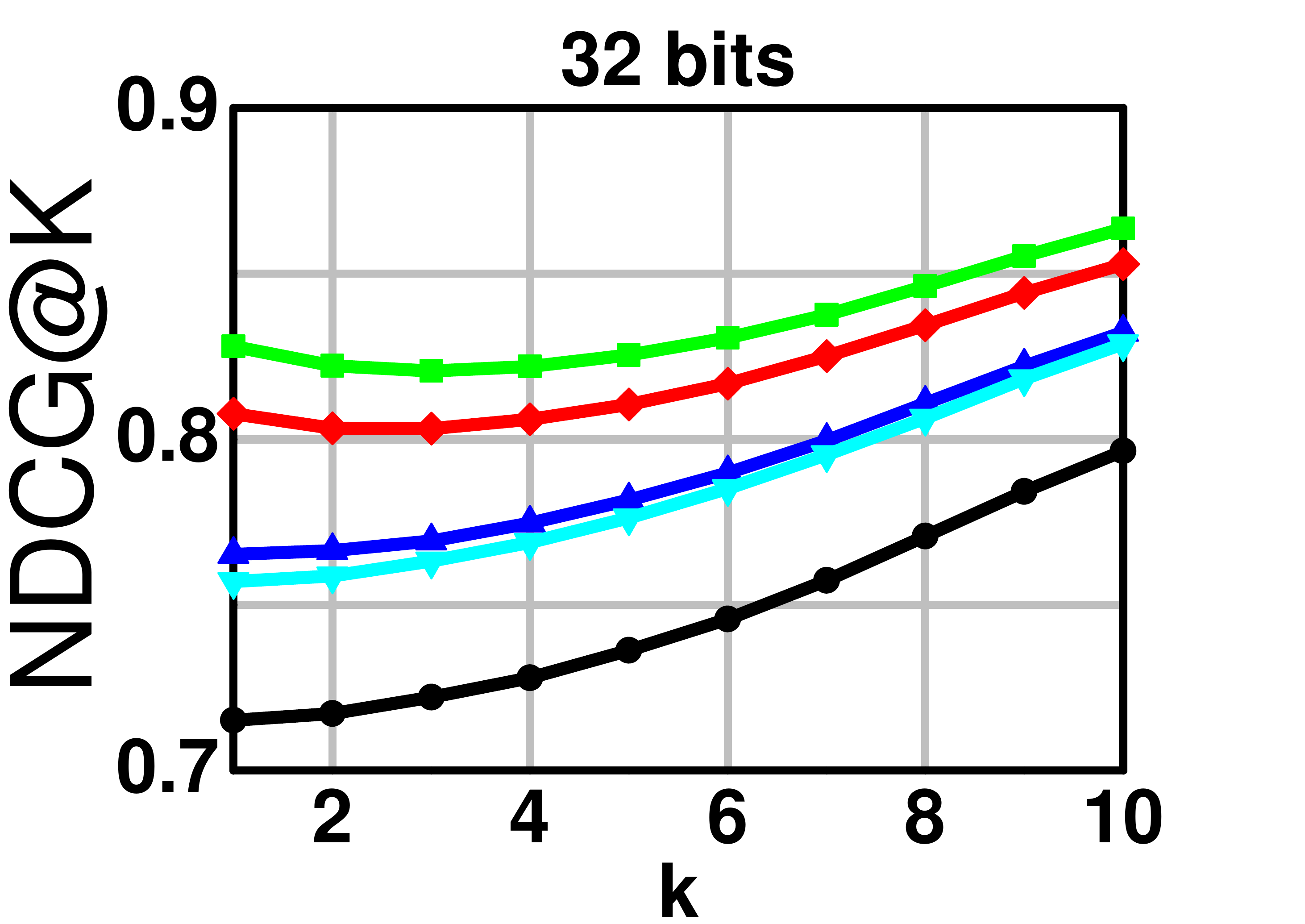}
\hspace{-0.24in}
\includegraphics[width=0.265\textwidth]{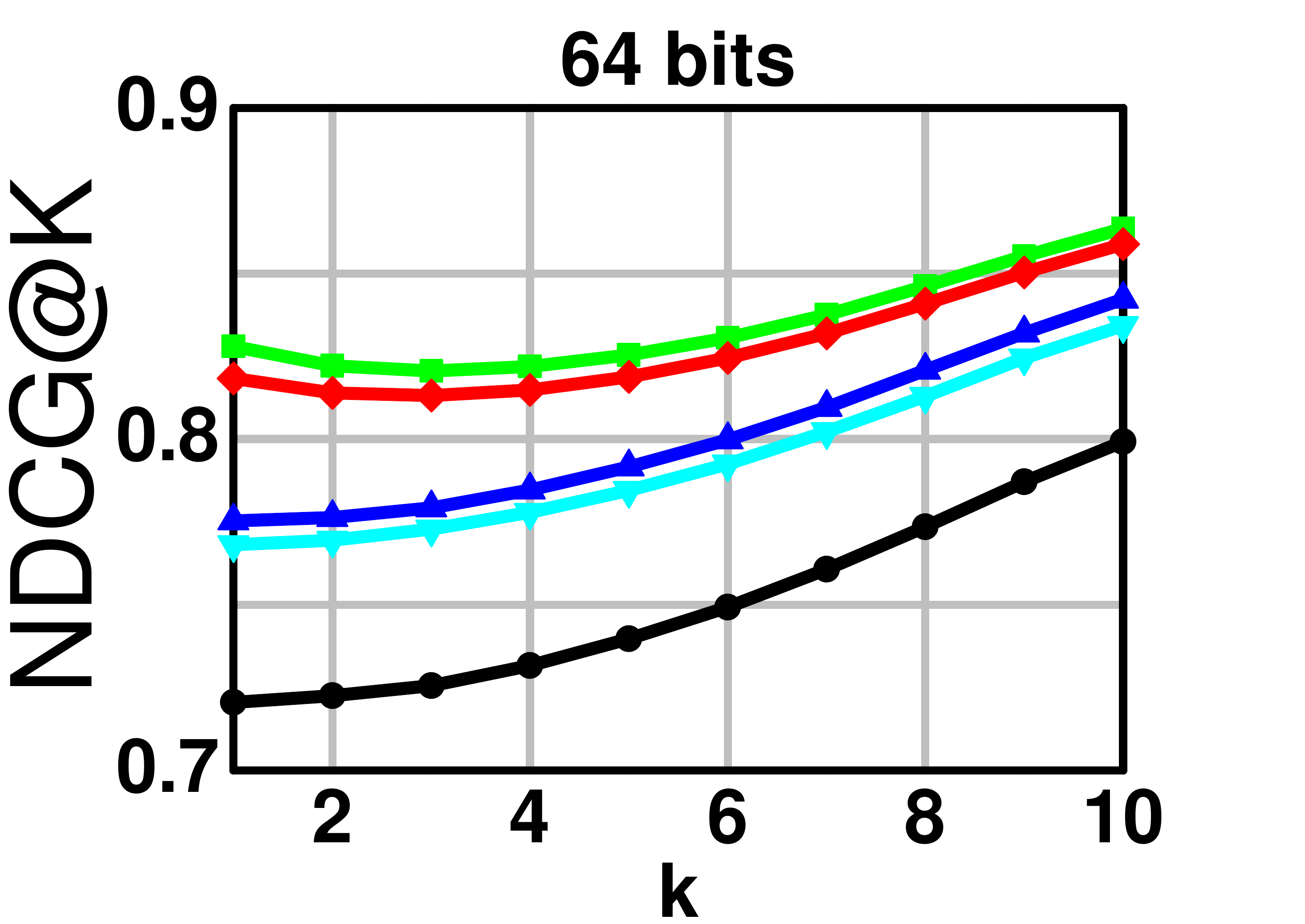}
\vspace{-0.2cm}
\caption{\textbf{Performance of NDCG@K (K ranges from 1 to 10) \textit{w.r.t.,} code length ranges for 8 to 64 on the two datasets.}}
\label{fig:performance}
\vspace{-0.3cm}
\end{figure*}
\subsection{Experimental Settings}
\textbf{Datasets}. We experiment on two publicly available datasets with explicit feedbacks from different real-world websites: \textit{Yelp} and \textit{Amazon}. Note that we assume each user has only one rating for an item and average the scores if an item has multiple ratings from the same user.

\textbf{a) Yelp.} This dataset \cite{Lian2017Discrete} originally contains 409,117 users, 85,539 items (points of interest on Yelp 
such as restaurants and hotels), and 2,685,066 ratings with integer scores ranging from 1 to 5. Besides, each item has a set of textual reviews posted by the users.

\textbf{b) Amazon.} This book rating dataset \cite{mcauley2015inferring} originally includes 12,886,488 ratings of 929,264 items (books on Amazon
from 2,588,991 users. In this dataset, an item also has a set of integer rating scores in $[1, 5]$ and a set of textual reviews.

Considering the extreme sparsity of the original Yelp and Amazon datasets,
we remove users with less than $20$ ratings and items rated by less than $20$ users. After the filtering, there are 13,679 users, 12,922 items, and 640,143 ratings left in the Yelp dataset. For the Amazon dataset, we remain 35,151 users, 33,195 items, and 1,732,060 ratings. For fair comparison with DCMF, we leave out the side information from the user field and represent an item with the bag-of-words encoding  of its textual contents after aggregating all review contents of the item. Note that we remove \textit{stopping words} and truncate the vocabulary by selecting the top 8,000 words regarding their \textit{Term Frequency–Inverse Document Frequency}. 
By concatenating the bag-of-words encoding (side information of the item) and one-hot encoding of user and item ID, we obtain a feature vector of dimensionality 34,601 and 76,346 for a rating (use-item pair) for Yelp and Amazon, respectively. \vspace{+5pt}
%

\noindent\textbf{Baselines}. We implement our proposed DFM method using Matlab\footnote{Codes are available: \href{https://github.com/hanliu95/DFM}{https://github.com/hanliu95/DFM}} and compare it with its real-valued version and state-of-the-art binarized methods for Collaborative Filtering:
\begin{itemize}[leftmargin=*]
\item \textbf{libFM}. This is the original   implementation\footnote{\href{http://www.libfm.org/}{http://www.libfm.org/}}
of FM which has achieved great performance for feature-based recommendation with explicit feedbacks. Note that we adopt $l_2$ regularization on the parameters to prevent overfitting and use the SGD learner to optimize it.
\item \textbf{DCF}. This is the first binarized CF method
that can directly optimize the binary codes~\cite{Zhang2016Discrete}.
\item \textbf{DCMF}. This is the state-of-the-art binarized method for CF with side information~\cite{Lian2017Discrete}. It extends \textbf{DCF} by encoding the side features as the constraints for user codes and item codes. 
\item \textbf{BCCF}. This is a two-stage binarized CF method~\cite{Zhou2012Learning} with a relaxation stage and a quantization stage. At these two stages, it successively solves MF with balanced code regularization and applies orthogonal rotation to obtain user codes and item codes.
\end{itemize}
Note that for \textbf{DCF} and \textbf{DCMF}, we use the original implementation as released by the authors. For BCCF, we re-implement it due to the unavailability. \vspace{+5pt}

\noindent\textbf{Evaluation Protocols}. We first randomly split the ratings from each user into training ($50\%$) and testing ($50\%$). 
As practical recommender systems typically recommend a list of items for a user, we rank the testing items of a user and evaluate the ranked list with \textit{Normalized Discounted Cumulative Gain} (NDCG), which has been widely used for evaluating ranking tasks like recommendation~\cite{NCF}. 
To evaluate the efficiency of \textbf{DFM} and real-valued FM, we use \textit{Testing Time Cost} (TTC) \cite{Zhang2016Discrete}, where a lower cost indicates better efficiency. \vspace{+5pt}

\noindent\textbf{Parameter Settings}. As we exactly follow the experimental settings of \cite{Lian2017Discrete}, we refer to their optimal settings for hyper-parameters of \textbf{DCMF}, \textbf{DCF}, and \textbf{BCCF}. For \textbf{libFM}, we test the $l_2$ regularization on feature embeddings $\mathbf{V}$ of $\{10^{-i} | i = -4, -3, -2, -1, 0, 1, 2\}$. Under the same range, we test the de-correlation constraint (\textit{i.e.,} $\beta$ in Eq. (\ref{eq:obj})) of \textbf{DFM}. Besides, we test the code length in the range of $[8, 16, 32, 64]$. It is worth mentioning that we conduct all the experiments on a computer equipped with an Intel(R) Core(TM) i7-7700k 4 cores CPU at 4.20GHZ, 32GB RAM, and 64-bit Windows 7 operating system.

\subsection{Performance Comparison (RQ1)}
In Figure \ref{fig:performance}, we show the recommendation performance (NDCG@1 to NDCG@10) of \textbf{DFM} and the baseline methods  on the two datasets. The code length varies from 8 to 64. From the figure, we have the following observations:
\begin{itemize}[leftmargin=*]
\item \textbf{DFM} demonstrates consistent improvements over state-of-the-art binarized recommendation methods across code lengths (the average improvement is 7.95\% and 2.38\% on Yelp and Amazon, respectively). The performance improvements are attributed to the benefits of learning binary codes for features and modeling their interactions.
\item Besides, \textbf{DFM} shows very competitive performance compared to \textbf{libFM}, its real-valued version, with an average performance drop of only 3.24\% and 2.40\% on the two datasets. By increasing the code length, the performance gap continuously shrinks from 5.68\% and 4.76\% to 1.46\% and 1.19\% on Yelp and Amazon, respectively. One possible reason is that \textbf{libFM} suffers from overfitting as the increase of its representative capability (\textit{i.e.,} larger code length)~\cite{NFM}, whereas binarizing the parameters can alleviate the overfitting problem. This finding again verifies the effectiveness of the proposed \textbf{DFM}.
\item Between baseline methods, \textbf{DCF} consistently outperforms \textbf{BCCF}, while slightly underperforms \textbf{DCMF} with an average performance decrease of 1.58\% and 0.76\% on the two datasets, respectively. This is consistent with the findings in \cite{Liu2014Collaborative} that the direct discrete optimization is stronger than two-stage approaches and that side information makes the user codes and item codes more representative, which can boost the performance of recommendation. However, the rather small performance gap between \textbf{DCF} and \textbf{DCMF} indicates that \textbf{DCMF} fails to make full use of  the side information. The main reason is because that \textbf{DCMF} performs prediction only based on user codes and item codes (which is same as \textbf{DCF}). This inevitably limits the representation ability of DCMF. 
\end{itemize}  

\begin{figure}
\centering
\includegraphics[width=0.252\textwidth]{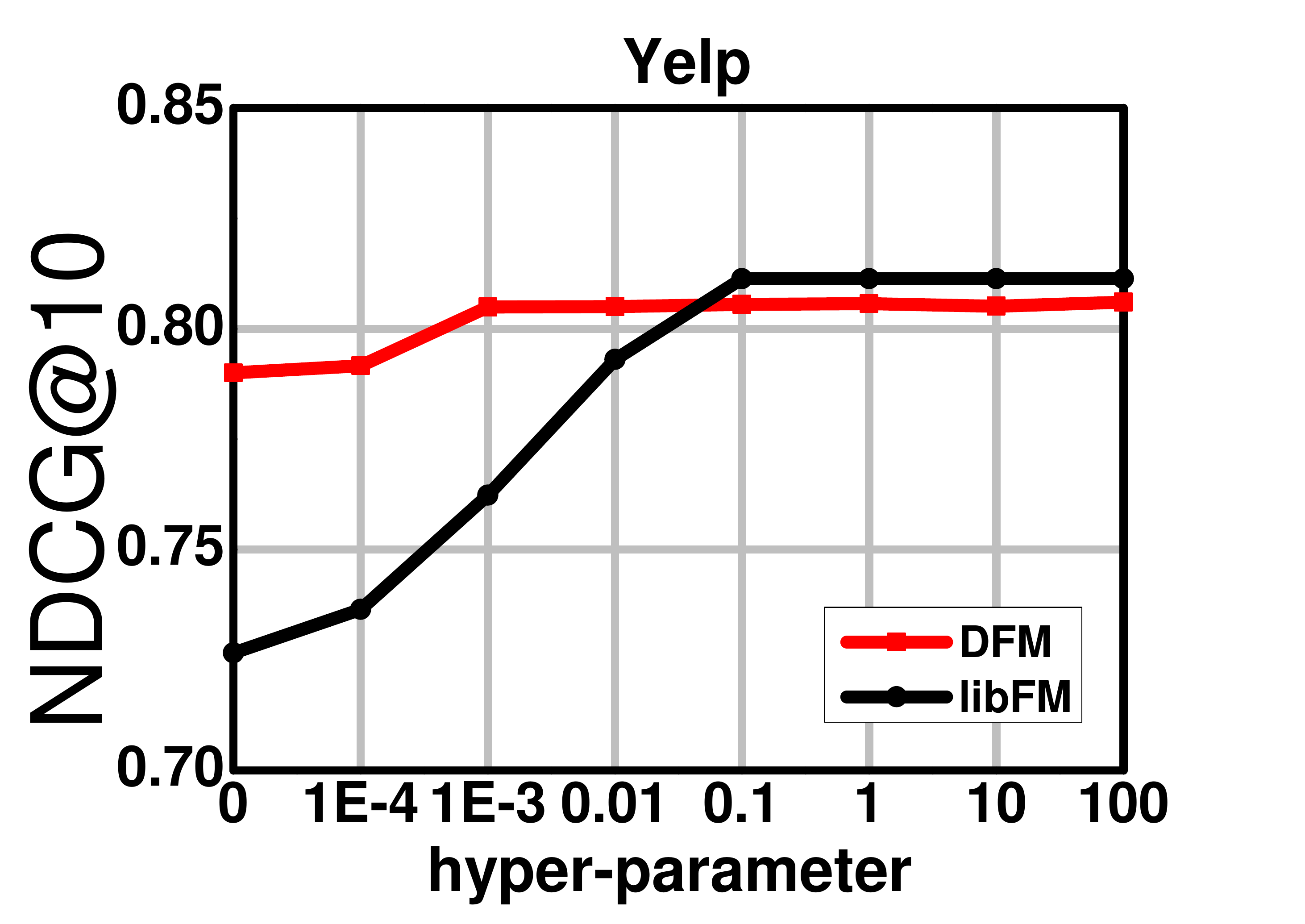}
\hspace{-0.24in}
\includegraphics[width=0.252\textwidth]{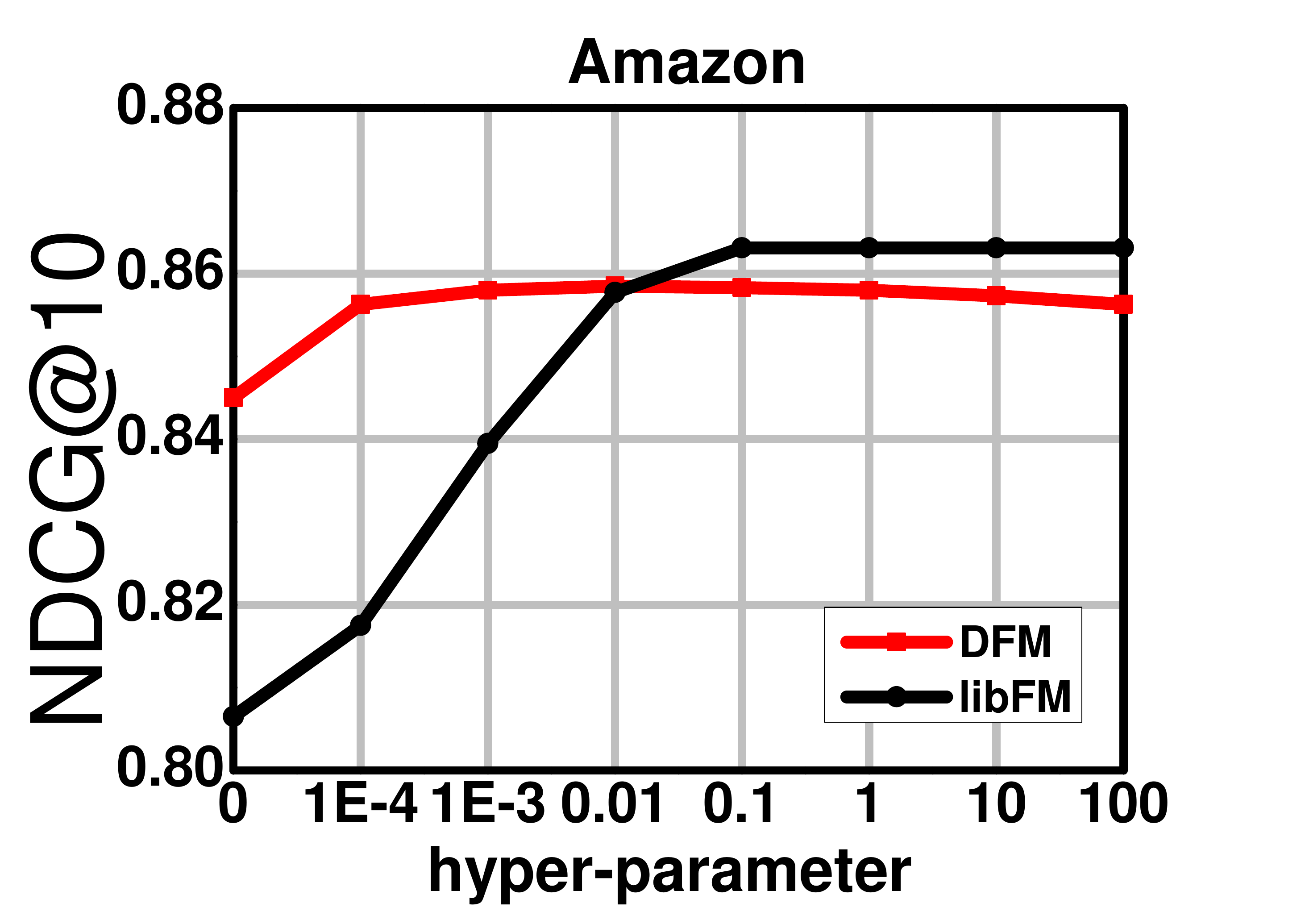}
\vspace{-15pt}
\caption{\textbf{Recommendation performance of libFM and DFM (code length=64) on NDCG@10 \textit{w.r.t.,} $l_2$ regularization (libFM) and de-correlation constraint (DFM).
}}
\label{fig:hyperparameter}
\vspace{-0.3cm}
\end{figure}

\subsection{Impact of Hyper-parameter (RQ2)}
Figure \ref{fig:hyperparameter} shows the recommendation performance of \textbf{libFM} and \textbf{DFM} on NDCG@10 regarding $l_2$ regularization of \textbf{libFM} and de-correlation constraint, respectively. We omit the results on different values of $K$ and code length other than $K = 10$ and code length = 64 since they shown the same trend. First, we can see that the performance of \textbf{libFM} continuously drops as we decrease the $l_2$ regularization. One reason is that \textbf{libFM} could easily suffer from overfitting \cite{xiao2017attentional}. Second, we observe that \textbf{DFM} performs slightly worse as decreasing the de-correlation constraint. By setting the de-correlation constraint and $l_2$ regularization to be zero, both of \textbf{DFM} and \textbf{libFM} exhibit significant performance decrease in NDCG@10. Specifically, the performance of \textbf{DFM} drops with a 1.91\% and 2.05\% margin on Yelp and Amazon, respectively, while \textbf{libFM} encounters a 10.44\% and 6.56\% one. The above findings again demonstrate the overfitting problem of \textbf{libFM}, which leads to \textbf{libFM} to be very sensitive to the $l_2$ regularization hyper-parameter, while the proposed \textbf{DFM} is relatively insensitive to its de-correlation constraint hyper-parameter.

\subsection{Efficiency Study (RQ3)}
As \textbf{libFM} is implemented based on C++, we re-implement the testing algorithm of \textbf{DFM} with C++ and compile it with the same C++ compiler (gcc-4.9.3) for a fair comparison. 
Table \ref{tab:efficiency} shows the efficiency comparison between \textbf{DFM} and \textbf{libFM} regarding TTC on the two datasets. We have the following observations:
\begin{itemize}[leftmargin=*]
\item \textbf{DFM} achieves significant speedups on both datasets regarding TTC, significantly accelerating the \textbf{libFM} by a large amplitude (on average, the acceleration ratio over \textbf{libFM} is 15.99 and 16.04, respectively). This demonstrates the great advantage of binarizing the real-valued parameters of FM.
\item The acceleration ratio of \textbf{DFM} based on \textbf{libFM} is stable around 16 times on both the datasets when the code length increases from 8 to 64. 
\end{itemize}
Along with the comparable recommendation performance of \textbf{DFM} and \textbf{libFM}, the above findings indicate that \textbf{DFM} is an operable solution for many large-scale Web services, such as Facebook, Instagram, and Youtube, to substantially reduce the computation cost of their recommendation systems.

\begin{table}[t]
  \centering
  \caption{\textbf{Efficiency comparison between DFM (C++ implementation) and libFM \textit{w.r.t.,} TTC (minutes) where the code length ranges from 8 to 64 on the two datasets.}}
  \vspace{-0.3cm}
  \textbf{Yelp}\\
  \vspace{+1pt}
  \resizebox{0.48\textwidth}{!}{
  \begin{tabular}{|c||c|c|c|c|}
  \hline
  \textbf{Code Length} &  \textbf{8}  & \textbf{16} &  \textbf{32} &  \textbf{64}\\
  \hline
  \textbf{libFM} (TTC) &$27.18$ & $56.77$ & $114.10$ & $217.64$ \\
  \hline
  \textbf{DFM} (TTC) &$2.06$ & $3.56$ & $6.60$ & $12.43$\\
  \hline
  Acceleration Ratio &$13.19$ & $15.95$ & $17.29$ & $17.51$\\
  \hline
  \end{tabular}
  }
  ~\\
  \vspace{+1pt}
  \textbf{Amazon}\\
  \vspace{+2pt}
  \resizebox{0.48\textwidth}{!}{
  \begin{tabular}{|c||c|c|c|c|}
  \hline
  Code Length &  8  &16 &  32 &  64\\
  \hline
  \textbf{libFM} (TTC) & $177.03$ & $357.46$ & $716.83$ & $1,414.67$ \\
  \hline
  \textbf{DFM} (TTC) &$12.67$ & $22.50$ & $42.56$ & $81.04$\\
  \hline
  Acceleration Ratio &$13.97$ & $15.89$ & $16.84$ & $17.46$\\
  \hline
  \end{tabular}
  }
  \label{tab:efficiency}
  \vspace{-0.3cm}
\end{table}
\section{Conclusions}
In this paper, we presented DFM, the first binary representation learning method for generic feature-based recommendation. 
In contrast to existing hash-based recommendation methods that can only learn binary codes for users and items, our DFM is capable of learning a vector of binary codes for each feature. As a benefit of such a compact binarized model, the predictions of DFM can be done efficiently in the binary space. Through extensive experiments on two real-world datasets, we demonstrate that DFM outperforms state-of-the-art hash-based recommender systems by a large margin, and achieves a recommendation accuracy rather close to that of the original real-valued FM. 

This work moves the first step towards developing efficient and compact recommender models, which are particularly useful for large-scale and resource-limited scenarios. 
In future, we will explore the potential of DFM for context-aware recommendation in mobile devices, a typical application scenario that requires fast and compact models. Moreover, we will develop pairwise learning method for DFM, which might be more suitable for personalized ranking task. 
With the fast developments of neural recommendation methods recently~\cite{NFM}, we will develop binarized neural recommender models in the next step to further boost the performance of hash-based recommendation. 
Besides, we are interested in deploying DFM for online recommendation scenarios, and explore how to integrate bandit-based and reinforcement learning strategies into DFM.
Lastly, we will explore the potential of DFM in other tasks such as 
popularity prediction of online content~\cite{feng2018learning}.
\vspace{+5pt} 

\noindent\textbf{Acknowledgment}
This work is supported by the National Basic Research Program of China (973 Program), No.: 2015CB352502; National Natural Science Foundation of China, No.: 61772310, No.: 61702300, and No.: 61702302; and the Project of Thousand Youth Talents 2016. This work is also part of NExT research, supported by the National Research Foundation, Prime Minister's Office, Singapore under its IRC@SG Funding Initiative.




\bibliographystyle{named}
\bibliography{ijcai18}

\end{document}